\documentclass[twocolumn,showpacs,pre,floatfix]{revtex4}
\usepackage{graphicx,amsmath}
\usepackage{txfonts}
\usepackage{color}

\begin{document}
\title{Threshold model of cascades in temporal networks}
\author{Fariba Karimi}
\affiliation{IceLab, Department of Physics, Ume{\aa} University, 90187 Ume\aa, Sweden}
\author{Petter Holme}
\affiliation{IceLab, Department of Physics, Ume{\aa} University, 90187 Ume\aa, Sweden}
\affiliation{Department of Energy Science, Sungkyunkwan University, Suwon 440-746, Korea}
\affiliation{Department of Sociology, Stockholm University, 10691 Stockholm, Sweden}
\begin{abstract}
Threshold models try to explain the consequences of social influence like
the spread of fads and opinions. Along with models of epidemics,
they constitute a major theoretical framework of social spreading
processes. In threshold models on static networks, an individual changes
her state if a certain fraction of her neighbors has done the same.
When there are strong correlations in the temporal aspects of contact
patterns, it is useful to represent the system as a temporal network.
In such a system, not only contacts but also the time of the contacts are represented explicitly. There is a consensus that
bursty temporal patterns slow down disease spreading. However, as
we will see, this is not a universal truth for threshold models. In
this work, we propose an extension of Watts' classic threshold model
to temporal networks. We do this by assuming that an agent is influenced
by contacts which lie a certain time into the past. I.e., the individuals are
affected by contacts within a time window. In addition to thresholds
as the fraction of contacts, we also investigate the number of contacts
within the time window as a basis for influence. To elucidate the
model's behavior, we run the model on real and randomized empirical
contact datasets.
\end{abstract}
\maketitle

\section{Introduction}

An important socio-economic mechanism is social influence---the spread
of opinions and beliefs from the social surrounding to an individual.
This type of process can be modeled as a threshold process---if the
fraction of neighbors in a static network exceeds a threshold, then
the focal vertex changes state. In his pioneering work from the 1970's
Mark Granovetter~\cite{granovetter_threshold} pointed out that social
collective behavior could be divided into processes depending on credulity
(including threshold models) or vulnerability (like disease-spreading
models). A recent, theoretically important development was made in
Ref.~\cite{watts_threshold}, where Duncan Watts proposed a threshold
model that is analytically tractable on networks. The study highlights
effect of network structure on cascade sizes. Some other studies have
tried to bridge the two classes---compartmental models and threshold
models~\cite{dodds_contagion_2004}. One of the key findings of Watts
was that network structure affects cascade processes~\cite{watts_threshold,centola}.
Another dimension that has been shown to be important for social spreading
phenomena is the timing of contacts~\cite{karsai_slow}. For this reason
it makes sense to model social contact patterns as temporal networks.
This is a term for network representations where the explicit time
of contacts is included. In this paper we extend Watts' cascade model to
temporal networks.

Watts' cascade model assumes that an actor can switch between two states.
It is a deterministic, non-equilibrium model of a cascade where actors
can change to a new state but not back to the previous one. In the original
model an actor is continuously influenced by its surrounding. In contrast,
in a temporal network, there is no fixed surrounding in time. One
has to decide what time into the past an actor can be influenced
by its contacts. In our model, we integrate the influence from contacts
over a time window. Our model is thus both taking into account the
chronological order of events and the period a contact that can influence
an actor. Outside the time window, communication does not influence
individuals. This is to say that for social influence, communications
in the past can be forgotten or become unimportant as time passes by.

We will use empirical temporal-network data as a substrate to run
our model on. We compare our results to those of randomized versions
of the original data. The results are compared to static and temporal
structural measures characterizing the temporal network.

\section{Methods}

%\subsection{Preliminaries}

The type of data we consider in this paper are sets of triples $(i,j,t)$,
or \emph{contacts}, which means vertices $i$ and $j$ have been in
contact at time $t$. Let $V$ be the set of $N$ vertices, $E$ the
set of $M$ edges (pairs of vertices that occur in at least one contact),
and let $C$ be the set of all $\Gamma$ contacts.

%\subsection{The threshold model of cascade in temporal network with time window }

We assume a system of vertices with one-to-one communication over edges
(pairs of vertices connected at least at one point in time). The interaction
is considered bidirectional in the sense that both vertices in contact
can influence one another. Each edge has a list of time stamps represent the time of communications. We assign a \emph{state} to each vertex. The state represents the thing which spread into the network and it is binary 0 or 1. State 0 corresponds to unchanged vertices or \emph{non-adopters}. State 1 corresponds to \emph{adopters}.The
population is initialized in the state 0 except one random vertex that is assigned
 to state 1. We study the spread of state 1 in networks. The term adopters comes from that early threshold models
often sought to capture the spread of new technology~\cite{valente_adopters}.

When we simulate the model, we follow the set of contacts in time order
and let each contact be an opportunity for the vertices to change state.
Vertices are influenced by their contacts within a finite \emph{time
window} from time $\theta$ in the past to the present. Let $f_{i}$
be the fraction of contacts between $i$ neighbors of state 1 within
the time window. In \emph{fractional-threshold model} if $f_{i}\geq\phi$ the vertex $i$ will be
in state 1 (for the remainder of the simulation). The reason that we choose the agent not to recover the 0-sate is to conform Duncan Watts' model, where he sought a maximally simple
model (to make it analytically tractable). One can compare this to disease
spreading models where the SI and SIR models do not allow the agent to
go back to its original state.
We also consider another version of the threshold where we consider absolute number $F_{i}$,
not the fraction, of interactions with state-1 neighbors and we change
the state if $F_{i}\geq\Phi$~\cite{balough_bootstrap,fontes_bootstrap} (we call this \emph{absolute-threshold model} as opposed to the previous fractional-threshold model).

\begin{figure}
  \includegraphics[width=\linewidth]{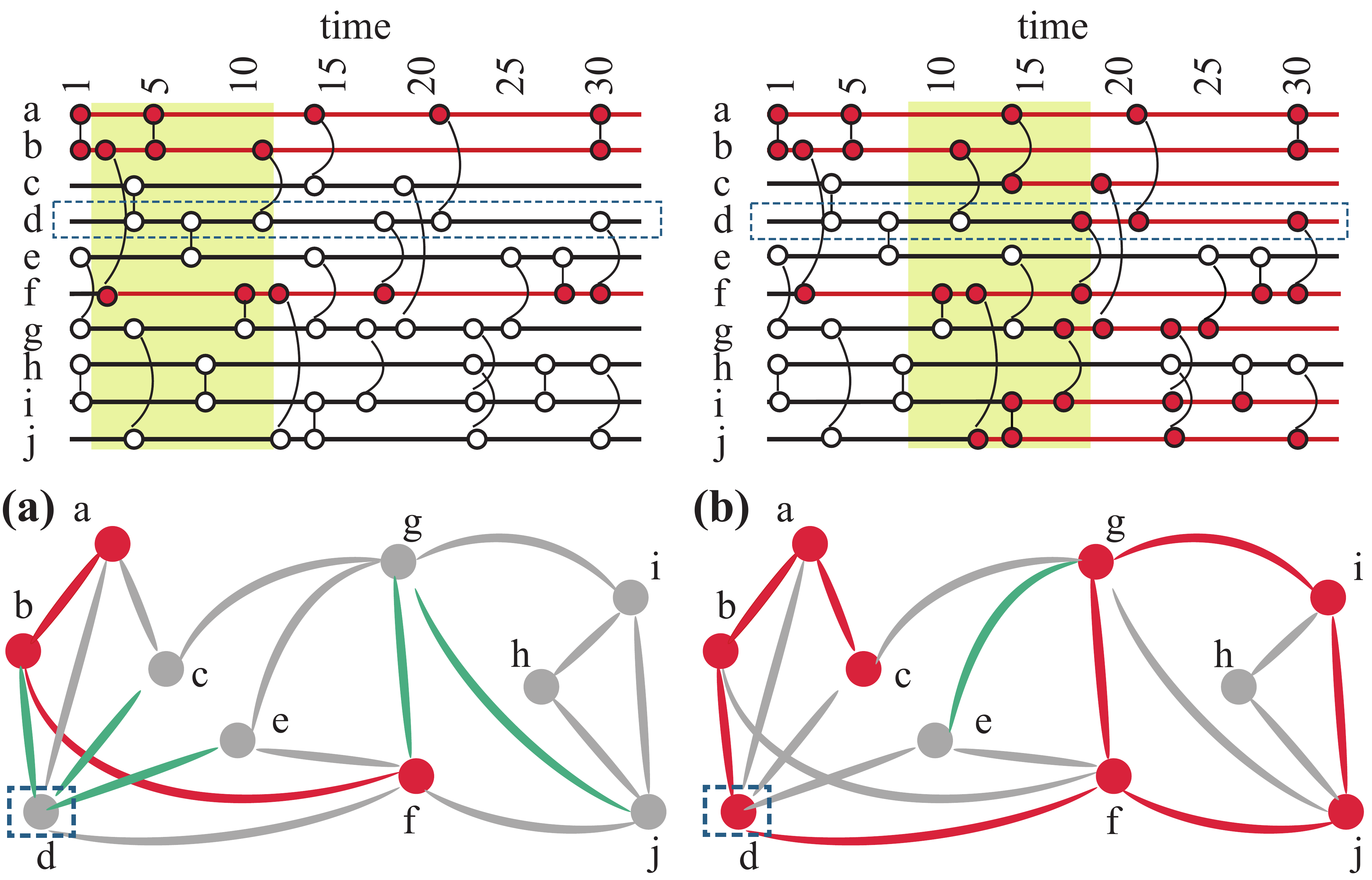}
  \caption{(Color online) An illustration of our temporal-network
cascade model. In the panel (a), vertex \emph{d} does not change its state
according to the time window and value of threshold. In panel (b),
vertex \emph{d} reaches the criteria to change its state.}\label{fig:cascade-illustration}
\end{figure}

Fig.~\ref{fig:cascade-illustration} illustrates the model. We use time line to represent the temporal network. Contacts
between individuals are illustrated as arcs in the time line. Red circles indicate adopters, white or gray circles indicate non-adopters. The threshold here is assumed to be $\phi=0.5$ and the time window $\theta=10$.
As time progresses, the time window slides through each contact and
updates vertices with respect to the contacts within the time window.
In panel (a), vertex \emph{d} does not change its state even though
it is in contact with red dot. At another time,see panel (b), vertex
\emph{d} changes its color because, inside the current time window,
the fraction of red neighbors exceeds the threshold.

We simulate the model on six empirical datasets. The results we will present are averaged over at least 100 runs.

\section{Empirical datasets}

We test our model on six empirical datasets generated by different
types of human interactions. The datasets were obtained with all individuals anonymized
to protect their identity. The first dataset consists of self-reported
sexual contacts from a Brazilian online forum where sex-buyers rate
and discuss female sex-sellers~\cite{rocha_epidemic}. The second
dataset comes from email exchange at a university~\cite{ekman_email}.
It was used in Ref.~\cite{barabasi_burst_2005} to argue that
human behavior often comes in bursts. The third datasets was collected
at a three-days conference from face-to-face interactions between
conference attendees~\cite{ht_contact_data}. The fourth dataset comes
from a Swedish Internet dating site where the interaction ranges from
partner seeking to friendship oriented~\cite{puss_och_kram}. The
fifth and sixth datasets come from a Swedish forum for rating and
discussing films~\cite{filmtipset_geman}. One of these datasets represents
comments in a forum that is organized so that one can see who comments
on whom. The other datasets comes from email-like messages. Table~\ref{tab:Properties-of-datasets} summarizes details of the datasets such
as number of vertices, number of contacts, sampling time and time
resolution. Some of the datasets like the movie forum, the email and the conference
contacts can be underlying structure for social influence, spread
of fads and ideas, or computer viruses. The Sexual-contact datasets and perhaps also
the online dating datasets, one can argue, represent the structure over
which sexually transmitted infections spread.

\begin{table*}
\caption{\label{tab:Properties-of-datasets}Summary of properties of the datasets.}
\begin{ruledtabular}
\begin{tabular}{c|cccc}
Data & No.\ vertices & No.\ contacts & time duration & resolution\tabularnewline \hline
Prostitution & 16730 & 50632 & 2232 days & day\tabularnewline
Email & 3188 & 309125 & 82 days & second\tabularnewline
Conference &  113 & 20,818 & 3 days & 20 seconds\tabularnewline
Online dating & 29341 & 536,276  & 512 days & second\tabularnewline
Internet community forum  & 6296 & 1,297,391  & 7 years & second\tabularnewline
Internet community messages & 35564 & 490,866  & 8 years & second\tabularnewline
\end{tabular}
\end{ruledtabular}
\end{table*}

\section{Fractional-threshold model}

\subsection{Effect of threshold values and time windows on cascade size}

\begin{figure*}
  \includegraphics[width=0.75\linewidth]{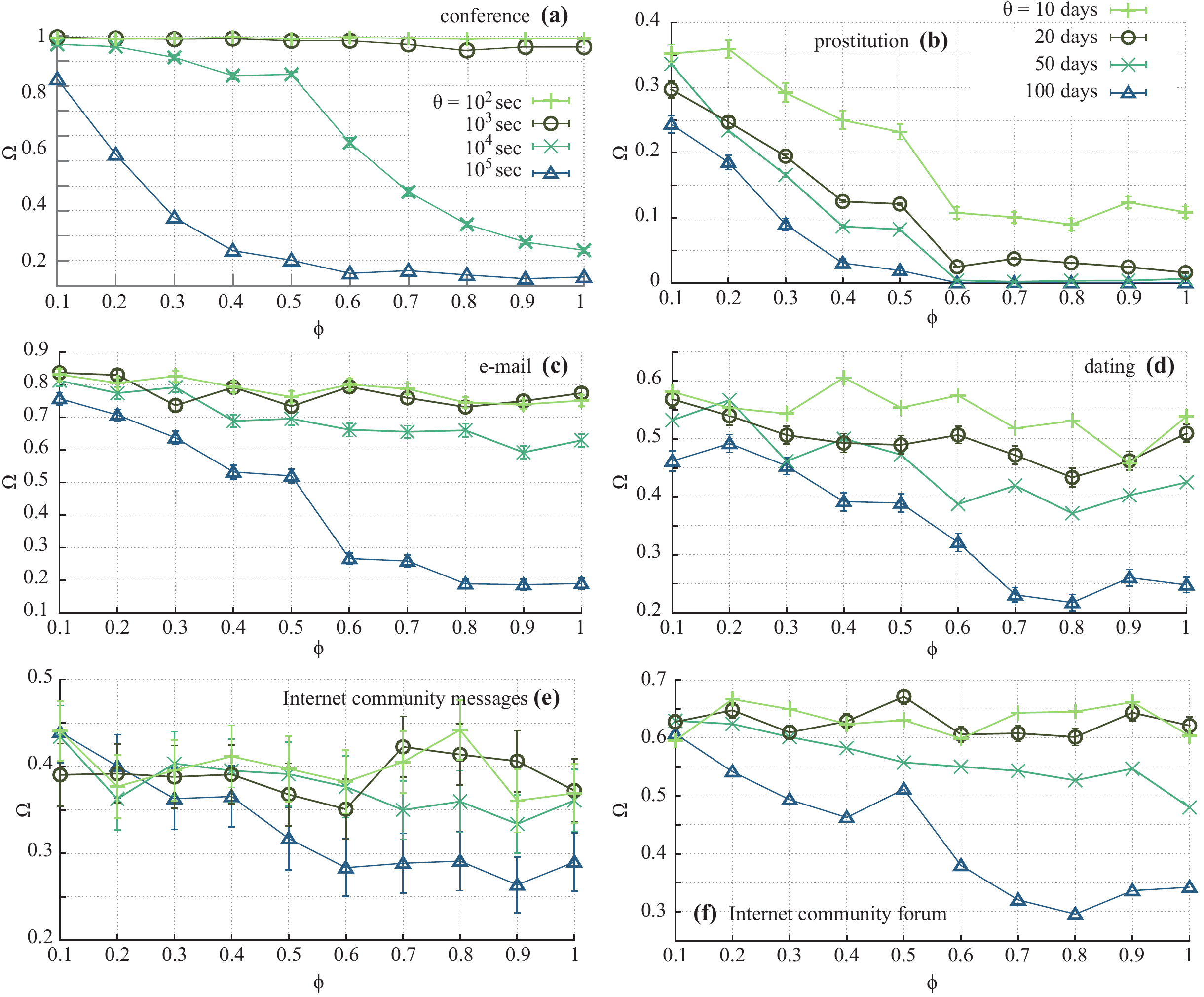}
  \caption{(Color online) Cascade size $\Omega$ versus threshold values $\phi$
from 0.1 to 1.0 for various time window sizes $\theta$. The symbols in panels (c)--(f) are the same as in (a). The error bars indicate the standard error over 200 runs of the cascade simulations.}\label{fig:cascade-threshold}
\end{figure*}

We start by investigating how the threshold value $\phi$ affects
the cascade size $\Omega$ for fixed values of time windows $\theta$. We define the cascade as a fraction of adopters over the whole population at the end of the simulation. As $\phi$
increases, agents naturally become more resistant to changing their
states. However, here we investigate how changing the time window
can influence $\Omega$ with respect to the threshold value. We test
our model in the mentioned empirical datasets by varying $\phi$ for
different values of $\theta$ (see Fig.~\ref{fig:cascade-threshold}).
Here we tune the threshold values and measure the size of cascades
as the fraction of adopters at the end of the simulation. The first
observation is that $\Omega$ varies
much from one datasets to another. It is hard to compare the actual
values of the different datasets since they have different time resolution
and sizes. The trends, however, are the same---the cascade sizes decrease
with $\phi$ for all $\theta$ and all datasets. We can understand that $\Omega$ decrease with $\theta$
since increasing the time window increases the expected
number of contacts, and for more contacts (when there are rather few
state-1 vertices), $f_{i}$ will decrease which decreases the probability
of changing state. In addition, the results show that unlike the static-network
counterpart~\cite{watts_threshold}, where cascade cannot trigger
for larger threshold values by a single seed, the time window makes it
possible for a cascade to propagate under larger threshold values
(because the number of contacts within a time window can be small
by fluctuations). This effect is also reflected by the finding that for the smallest
$\theta$-values the curves fall on top of each other. In these cases,
these are rarely more than one contact within the time window, so
that when there is a contact between a 0- and 1-individual the 0-individual
will always become adopter. In this case, the model becomes effectively
a disease-spreading model with 100\% transmission probability.

\subsection{Effect of temporal structure on cascade size}

Now we will turn to a more direct measurement of the effects of temporal-network
structure on the cascade dynamics. Studies have been shown that human
activities often come in bursts~\cite{barabasi_burst_2005,goh-burst} meaning that high intensity of activites in short interevent time followed by long interevent time with low activity. 
The burstiness can influence spreading phenomena over the contacts~\cite{vazquez}.
One straightforward way to detect the effect of temporal structure
on cascade dynamics is to compare the results with a null model. We
follow Ref.~\cite{karsai_slow} and use a null model derived by randomly
permuting the timing of contacts and keeping the network structure and the
number of contacts per vertex unchanged. This randomization keeps
human daily patterns intact but destroys effects of the order of events.

\begin{figure*}
  \includegraphics[width=0.8\linewidth]{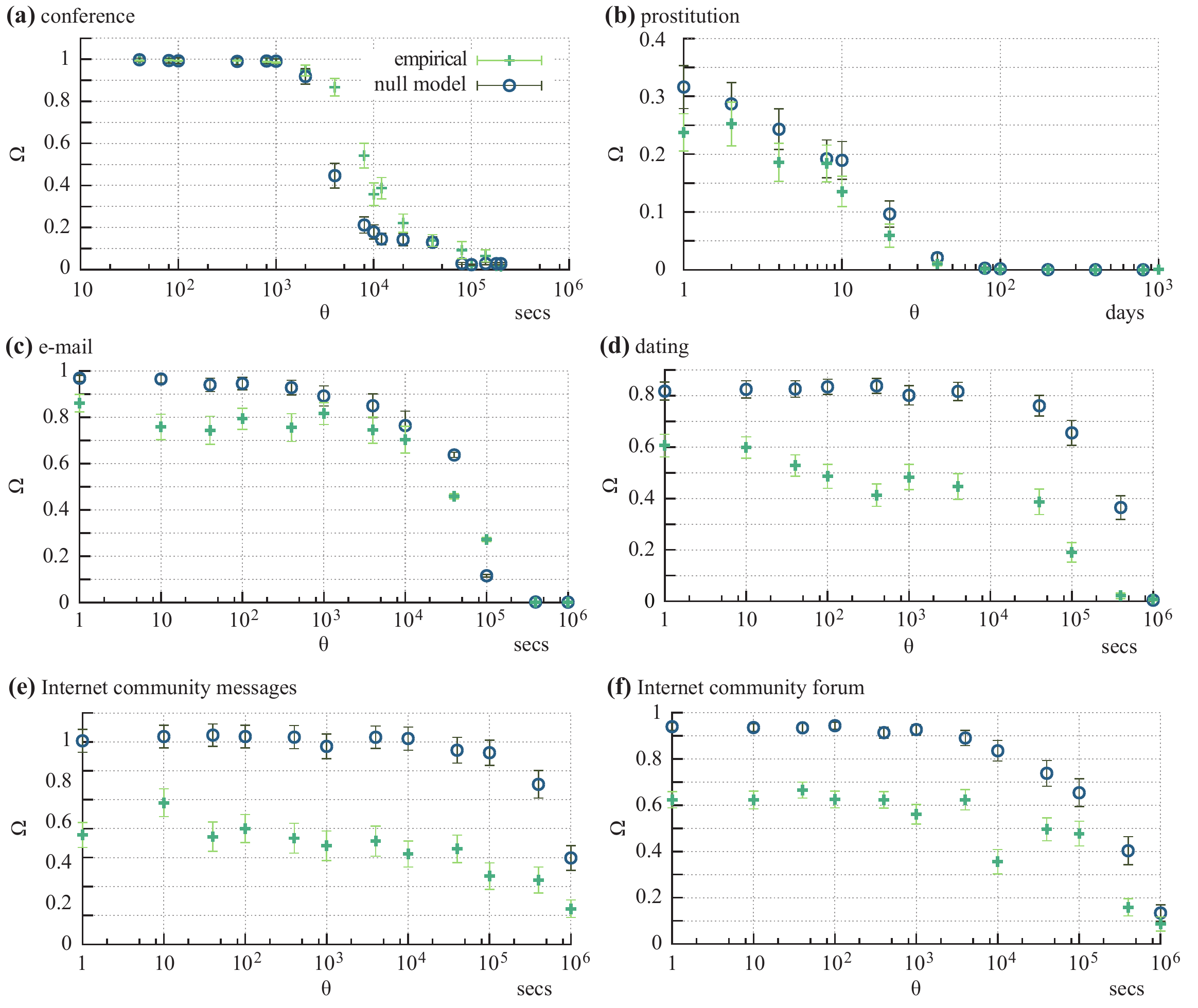}
  \caption{(Color online) Cascade size versus time-window
size for fixed value of the threshold ($\phi=0.7$). To verify the
effect of temporal correlations, we compare the data to a null model
by reshuffling time stamps. The figures show that, except the conference
data, temporal correlations slow down the cascade size compare to
the randomized networks without temporal structure. The error bars indicate the standard error over 150 runs of cascade simulations.}\label{fig:Cascade-size-versus-time}
\end{figure*}

Fig.~\ref{fig:Cascade-size-versus-time} shows the results of changing
the size of the time window on cascade size for six empirical datasets.  The null model is constructed by keeping the edges and degree fixed and reshuffling time stamps. In such model, the topological aspect of the networks preserved and only the temporal structure is reshuffled. Thus, the null model behaves similar to the real data except the effect of temporal structure.
We set the threshold value to $\phi=0.7$ for two reasons. First,
for this value the system responds more strongly to changes in $\theta$.
Second, this is a region where the threshold model on static
network does not give any cascades. One observation is that $\Omega$ are all decreasing as a function of the time-window size. This
decay is (inverse) sigmoidal with a plateau, followed by a rapid decay
to zero. The fact that the cascade sizes are zero for large $\theta$ is not surprising since as the time window increases, it is
less likely for cascade to happen in high threshold value due to an increase
in the number of contacts. In short, the temporal-network version of the
cascade model becomes more similar to what happens on a static network (when $\theta$ is sufficiently large),
and there we know that large threshold values do not support cascades.
	
For all datasets, except the conference data, the null-model has larger
cascades. For the conference data, this is interesting because it has been observed in disease
spreading models that the time correlation slows down the spreading~\cite{karsai_slow}. If we assume that faster spreading corresponds to larger cascades then we see the opposite effect
in the conference data. We also see that the online dating and Internet community data
are the ones with largest differences between the real data and the null models. The connectance in these datasets is low, assortativity is neutral and burstiness is high (see Table~\ref{tab:properties-networks}). Note that we perform the measurments in Table~\ref{tab:properties-networks} in static, aggregated networks.

The prostitution dataset has the low connectance and lowest value of assortativity compare to null model (see Table~\ref{tab:properties-networks}). This may thus be a case when the network topology affects the system's sensitivity to the temporal structure. One possible reason could be
that assortativity and burstiness are positively correlated---see Table~\ref{tab:properties-networks}
where the prostitution and email datasets with lower assortativity also have
lower burstiness. We will not go deeper in this analysis in this
paper but take it as a challenge for the future.

For the e-mail and prostitution data the null model and empirical data behave similarly. The reason we believe is the interplay between connectance, assortativity and burstiness. Higher connectance in the e-mail and lower burstiness in the prostitution can cause the coincide effect.

\begin{figure}
\includegraphics[width=0.9\linewidth]{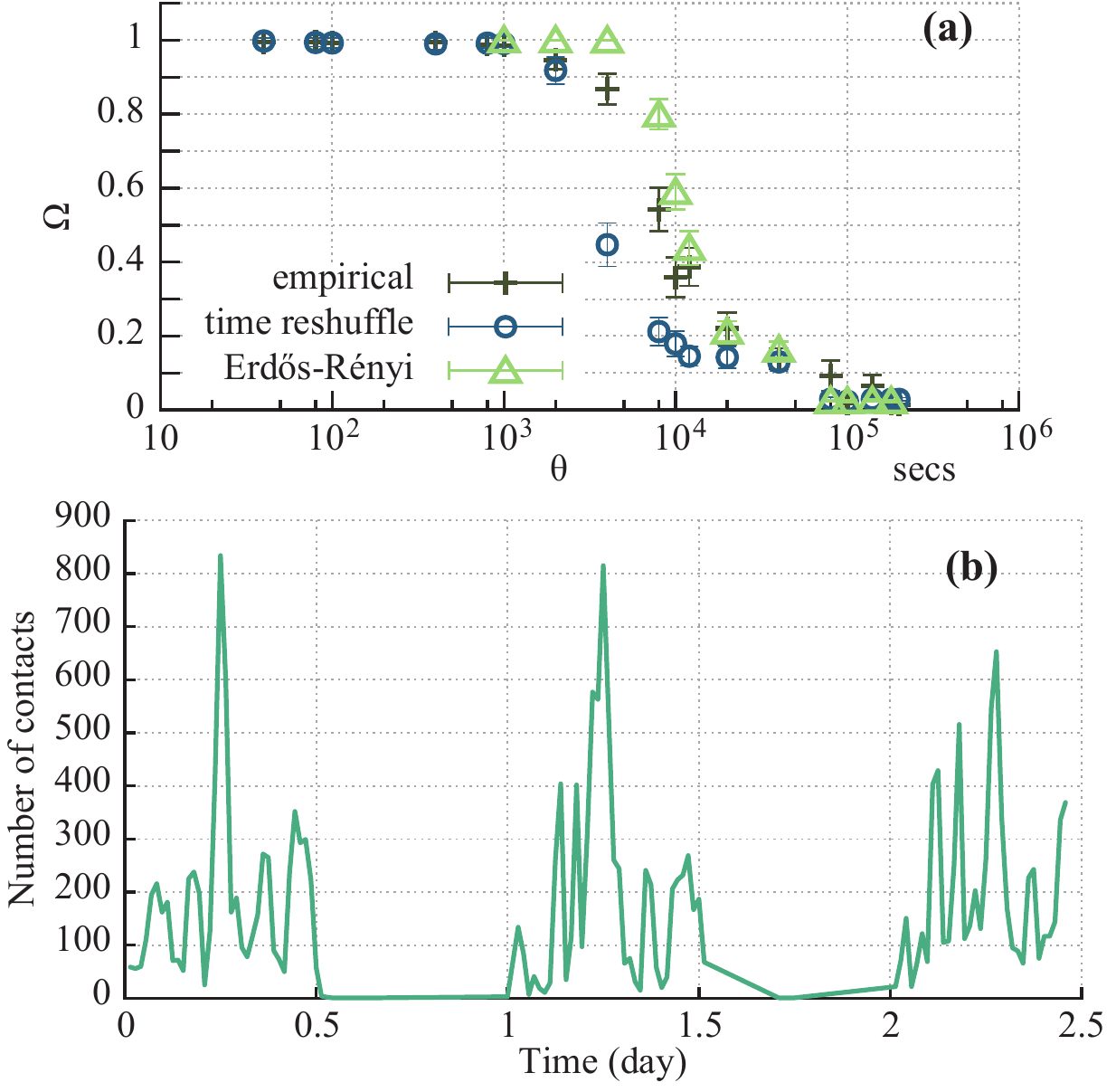}
\caption{\label{fig:conference-compare} (Color online.)  Panel (a) shows the cascade size $\Omega$ as a function of time-window size $\theta$ for the conference data. The results are compared with null
model by reshuffling time stamps and generated network based on the
Erd\H{o}s-R\'{e}nyi model with the same time-stamps as in the empirical
data. The curve for the real data lies somewhere between Erd\H{o}s-R\'{e}nyi
model with no specific network structure and null model with reshuffled
time stamps where burstiness and temporal effect is gone. The error bars indicate the standard error over 150 runs of cascade simulations. (b) displays
contact frequencies over time in conference contact datasets illustrating the regular meeting
patterns of the conference. Peaks can be seen during coffee break and lunchtime.}
\end{figure}

In the conference dataset, the temporal structure makes the cascades larger than in the randomized model without a temporal structure. This behavior is different from the other dataset where randomized models have larger cascade size compare to real dataset. The difficulty in explaining this is that there are many differences between the conference data and the others. One of the differences
is the way burstiness is manifested. In this data, bursts occur since
people are organized to meet during coffee breaks and lunch (Fig.~\ref{fig:conference-compare}a).
So the burstiness is designed rather than self-organized by human
behavior. This also means that all people will be in bursts of activity
at the same time. Another property that sets the conference data aside
from the others is that it is denser. The connectance (fraction of all possible links that are realized~\cite{connectance} ) is almost 60
times larger than the second most densely connected dataset (see Table~\ref{tab:properties-networks}). This connects to the observation
above that networks of high assortativity have lower cascade sizes
for the randomized data. Typically, networks of high assortativity
have a densely connected subgraph---for the conference data the entire
network is such a densely connected graph. It is known that the assortativity
is dependent on the connectance~\cite{holme-assortativity},
so the assortativity value for the conference data in Table~\ref{tab:properties-networks}
should be taken with a grain of salt.

To investigate why the conference dataset behaves differently compare
to the other datasets, we compare the cascade size as a function of
time-window size for the empirical data and two models. One of the
models is the Erd\H{o}s-R\'{e}nyi model with the same number of nodes and vertices as in
the original data. We assign the same time stamps to the edges as
in the real data (so that an edge has the same number of contacts
happening at the same times as an edge in the empirical data). This
model takes away all the network structure (except the size of the
network) and preserves the temporal correlation. The other model we
use takes away all the temporal structure but preserves the network
topology. Like above we shuffle the time stamps and keep the accumulated
network. We observe---in Fig.~\ref{fig:conference-compare}a---that the
temporal structure of the empirical data make the cascade bigger (as
observed above for intermediate time windows) and the network structure
make the cascades smaller. To explain this we go a bit further in
our description of the dataset. The conference contact data represents
face-to-face interaction recorded by radio frequency detectors. A special
feature, compared to the other datasets (except, to a smaller degree, the e-mail data), is that the method enables
to record simultaneous contacts. This together with the above observation
that the interaction at the conference is more organized (Fig.~\ref{fig:conference-compare}b)
and people meet more intensely during some times means that compared
to the other datasets (except the prostitution data that has a very
low time resolution) the conference data has a many of overlapping
contacts (where one actor is in contact with more than one other in single time unit).
To be precise, it has four times more such contacts than the one with
the second largest value. One consequence of this is that the $f_{i}$-value
will fluctuate more, and these fluctuations can promote the cascade.
That the Erd\H{o}s-R\'{e}nyi model and the real data are rather similar tells us that the cascade model is highly related to temporal structure of the network not the topology. As it can be seen in panel in Fig.~\ref{fig:conference-compare}a, the empirical and Erd\H{o}s-R\'{e}nyi behave similarly, while reshuffling time stamps, slow down the cascade. The conference network is very dense (Table~\ref{tab:properties-networks}),
lacking community structure (the whole network is a \emph{de facto} community
itself) or skewed degree distributions that could influence the spreading.

\begin{table*}
\begin{ruledtabular}
\begin{tabular}{c|cccc}
The data & connectance & burstiness & clustering coefficient & assortativity\tabularnewline
\hline
Prostitution & 0.0002 & 0.44 & 0.00 [0.00] & $-0.10$ $[-0.02]$\tabularnewline
Email &0.006 & 0.68 & 0.06 [0.07] & $-0.25$ $[-0.22]$\tabularnewline
Conference  & 0.34 & 0.74 & 0.50 [0.48] & $-0.12$ $[-0.17]$\tabularnewline
Online dating & 0.0002 & 0.65 & 0.00 [0.00] & $-0.04$ $[-0.04]$\tabularnewline
Internet community forum & 0.006 & 0.86 & 0.26 [0.21] & $-0.1747$ $[-0.1797]$ \tabularnewline
Internet community messages & 0.0001 & 0.61 & 0.05 [0.00] & $-0.0377$ $[-0.0440]$ \tabularnewline
\end{tabular}
\end{ruledtabular}
\caption{\label{tab:properties-networks}Properties of the aggregated (and hence static) networks. Connectance is the fraction of all pairs of vertices that are edges~\cite{connectance}. Burstiness
is the coefficient of variation of the inter-event times of the activities
of each vertex~\cite{goh-burst}. The clustering coefficient measures
the fraction of triangles relative to the number of connected triples~\cite{newman2010}. Assortativity measures the correlation of the degrees
at either side of an edge~\cite{newman2010}. Clustering coefficient and assortativity for null models are written in brackets.}
\end{table*}

\section{Absolute-threshold model}

The fractional-threshold model, described in the previous section, does
not cover all imaginable situations of social influence spreading
in reality. One can also imagine that an agent needs an absolute number
of influences during a time period to change state. In this section,
we investigate such a modification of the model. One can assume that
many real systems are in between these two versions of the threshold
models. We denote the absolute-value threshold by $\Phi$, and assume,
an agent changes state when the absolute number of contacts with an
adopter during the time window is $F_{i}=f_{i}c_{i}\geq\Phi$ (where
$c_{i}$ is the number of contacts of $i$ within the time window).

\begin{figure*}
  \includegraphics[width=0.8\linewidth]{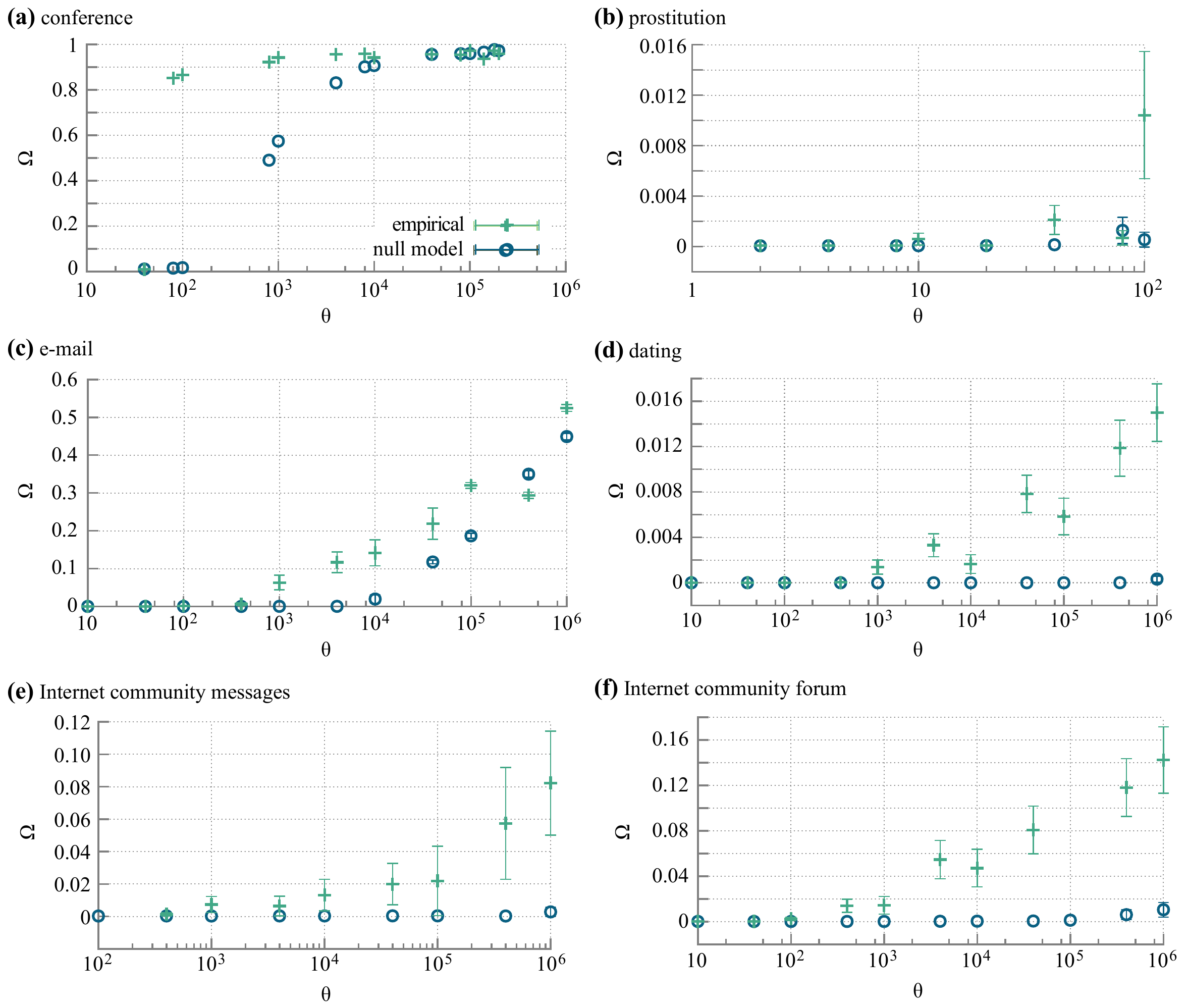}
  \caption{(Color online) Cascade size versus time windows
for the absolute-threshold model. The threshold is $\Phi=4$
(except the prostitution data where it is $\Phi=2$). To investigate
the effect of temporal correlations, we compare the data to a null
model where the times of the contacts are randomly shuffled. In most
cases, the temporal structure makes the cascades larger as time windows
increase. The error bars indicate the standard error over 150 runs of cascade simulations. }\label{fig:Cascade-size-absolute}
\end{figure*}

In Fig.~\ref{fig:Cascade-size-absolute}, we show the values of cascade
sizes for various time windows. Except for the prostitution data, $\Phi$ is fixed to 4. For
the prostitution data, we choose $\Phi=2$, because the average number
of contact per vertex are lower. A first observation is that the cascade sizes
increase with the time-window size. This is the opposite result than the
fractional-threshold model. This trend comes from the fact that an agent would
meet more adopters when exposed to more contacts due to the longer
time-window. For the fractional-threshold case this is balanced by
the increasing number of non-adopters the agent meets, but not so
in the absolute-threshold model. In
contrast to the previous results, the cascade-size for the
empirical data is larger than the null-model. Thus, the temporal
correlations in the fractional-threshold model (shown in Fig.~\ref{fig:cascade-threshold}) boost the possibility for a non-adopter
to meet adopters within the time window compared to contacts happening
in random order. 

For the conference data, the temporal correlation has a huge impact for intermediate time windows. For larger time windows, the temporal correlation does not play a role since the connectance is high and time window is large enough for cascade to trigger. For the datasets such as dating, Internet community messages and Internet community forums, the intermediate time windows has a large impact for cascade compare to null models. In these datasets, the sampling time is still larger compare to the time windows. We can see that, the effect of temporal correlation is larger for low connected networks. As time window increases, the cascade hardly can trigger in time-reshuffled models. In the e-mail dataset, the null model and the empirical data behave similarly. This behavior suggests that the relatively high connectance and the low burstiness compensate the temporal correlation on cascade size.

It would be interesting for the future to study the
intermediate case when agents are governed by a sub-linear threshold
function---i.e.\ that an agent changes state if $f_{i}c_{i}^{\alpha}\geq\Phi$
where $0\leq\alpha\leq1$---both because real social systems could possibly
belong to this region and that the extreme values are conspicuously
different.

\section{Discussion}

We have studied threshold models of cascades in temporal network by
extending Watts' cascade model~\cite{watts_threshold}. A key assumption
is that people are influenced by contacts dating some time back into
the past. We investigated two versions of the model, one where people
respond to a threshold in the fraction of adopters, one where they respond to the absolute number of such contacts. One observation
is that these systems are heavily affected by the temporal network
structure. This can be seen in that the two models respond differently to randomization
of the time stamps (meaning that one remove influence of the order
of events). Then there is a very large difference between the fractional-
and absolute-threshold models. In the former case, the cascade
sizes decrease with time-window size; in the latter case,
it is the other way around. In addition, the response to randomization is
different---for the fractional-threshold case the temporal-network structure
makes the cascades larger, whereas in the absolute-threshold
case, randomization makes cascades smaller. This is interesting in
the light of Ref.~\cite{karsai_slow} where the authors argue that
burstiness slows down spreading phenomena. The authors have disease-spreading
models in mind, but the conclusion seems not to generalize to threshold
models. This is assuming that one can identify large
outbreak size and with transmission speed, which in most usual situations probably is true.

One of the datasets---the data of who talks to whom during a conference---did
not fit completely to the above picture. For this dataset and the fractional-threshold
model the cascade size responded differently to randomization---the cascade
sizes decrease while those of the other datasets increase.
For the absolute-threshold model the effect of randomizations
is the same, but the shape of the curves is very different. We attribute
the former observation to the fact that this dataset is very densely
connected. The connectance (fraction of all possible links that are realized~\cite{connectance} ) is
over fifty times larger than any other dataset. The shape of the curves
can be explained as a similar effect---since there are so many contacts
per vertex, the cascades reach the entire systems once the time window
is large enough. Other datasets have comparatively low activity and
all the contacts are insufficient to trigger global cascades. Furthermore, the interplay between connectance and burstiness, has a significant impact on threshold cascade models. 

This is, we believe, only the beginning of the exploration of the temporal-network effects on threshold models. We hope future studies can connect how the interplay between network topology and temporal
correlations affect cascades in populations.

\acknowledgments{This research was the Swedish Research Council and the WCU program through NRF Korea funded by MEST R31-2008-000-10029-0. The authors thank Taro Takaguchi for comments.}

\bibliographystyle{apsrev}
\bibliography{cascade_references}

\end{document}